\documentclass[11pt]{article}
\pdfoutput=1	% needed to include PDF graphics

\usepackage[top=2cm, bottom=2cm, left=2cm, right=2cm]{geometry}
\usepackage{jcappub}
\usepackage{float}
\usepackage{graphicx}
\usepackage{rotating}
\usepackage{color}
\usepackage{amsmath,bm}

\newcommand{\sv}{\langle\sigma v \rangle}

\newcommand{\J}{J}
\newcommand{\diff}{\mathrm{d}}

\definecolor{plotpink}{RGB}{205,0,180}
\definecolor{plotcyan}{RGB}{0,215,215}
\definecolor{plotblue}{RGB}{0,0,235}
\definecolor{plotorange}{RGB}{245,140,0}
\definecolor{plotgreen}{RGB}{30,130,0}
\definecolor{plotred}{RGB}{240,0,0}
\definecolor{darkgreen}{RGB}{0,170,0}

\title{Probing dark matter annihilation in the Galaxy with antiprotons and gamma rays}
\author[a]{Alessandro Cuoco,}
\author[a]{Jan Heisig,}
\author[a,b,c]{Michael Korsmeier}
\author[a]{and Michael Kr\"amer}
\affiliation[a]{Institute for Theoretical Particle Physics and Cosmology,
RWTH Aachen University, 52056 Aachen, Germany}
\affiliation[b]{Dipartimento di Fisica, Universit\`a di Torino, via P. Giuria 1, 10125 Torino, Italy}
\affiliation[c]{Istituto Nazionale di Fisica Nucleare, Sezione di Torino, Via P. Giuria 1, 10125 Torino, Italy}

\emailAdd{cuoco@physik.rwth-aachen.de}
\emailAdd{heisig@physik.rwth-aachen.de}
\emailAdd{korsmeier@physik.rwth-aachen.de}
\emailAdd{mkraemer@physik.rwth-aachen.de}

\abstract{
A possible hint of dark matter annihilation has been found in Cuoco, Korsmeier and Kr\"amer (2017)
from an analysis of recent cosmic-ray antiproton data from AMS-02 and taking into account 
cosmic-ray propagation uncertainties by fitting at the same time dark matter and propagation parameters.
Here, we extend this analysis to a wider class of annihilation channels.
We find consistent hints of a dark matter signal with an annihilation cross-section close to 
the thermal value and with masses in range between 40 and 130\,GeV depending on the 
annihilation channel. Furthermore, we investigate in how far the possible signal is compatible 
with the Galactic center gamma-ray excess and recent observation of dwarf satellite galaxies 
by performing a joint global fit including uncertainties in the dark matter density profile. 
As an example, we interpret our results in the framework of the Higgs portal model.
}

\subheader{\hfill \textnormal{TTK-17-09}}
\keywords{}
\arxivnumber{}

\begin{document}

\maketitle
\flushbottom

%===================================================================
\section{Introduction}\label{sec:intro}
%===================================================================

Observations of cosmic-ray (CR) antiprotons are a sensitive probe of dark matter (DM) models with thermal annihilation cross sections~\cite{Bergstrom:1999jc,Donato:2003xg,Bringmann:2006im,Donato:2008jk,Fornengo:2013xda,Hooper:2014ysa,Pettorino:2014sua,Boudaud:2014qra,Cembranos:2014wza,Cirelli:2014lwa,Bringmann:2014lpa,Giesen:2015ufa,Jin:2015sqa,Evoli:2015vaa}. 
In particular, with the very accurate recent measurement of the CR antiproton flux 
by \mbox{AMS-02}~\cite{Aguilar:2016kjl}, 
it is a timely moment to investigate this subject.
A joint analysis of the CR fluxes of light nuclei and a potential DM contribution 
to the antiproton flux provides strong DM constraints~\cite{Cuoco:2016eej},
as well as an hint for a DM signal  corresponding to a DM mass of about 80\,GeV, and a thermal hadronic annihilation cross section, $\left\langle \sigma v \right\rangle \approx 3 \times 10^{-26}$~cm$^3$/s.  These values have been derived in  \cite{Cuoco:2016eej}
in a novel analysis where CR propagation uncertainties have been marginalized away,
taking into account possible degeneracies between CR uncertainties and DM.
A similar result has been found in~\cite{Cui:2016ppb}, but using the boron over carbon ratio, also recently measured by AMS-02~\cite{Aguilar:2016vqr}.
In the present work, we shall extend the DM analysis of the CR antiproton flux presented in~\cite{Cuoco:2016eej} to a comprehensive set of standard model (SM) annihilation channels, 
including gluons, bottom quarks, $W$, $Z$ and Higgs bosons, as well as top quarks.

Similarly, an excess in gamma-ray emission toward the center of our Galaxy has been reported by several analyses~\cite{Vitale:2009hr,Goodenough:2009gk,Hooper:2010mq,Hooper:2011ti,Abazajian:2012pn,Hooper:2013rwa,Gordon:2013vta,Abazajian:2014fta,Daylan:2014rsa,Calore:2014xka,TheFermi-LAT:2015kwa,Karwin:2016tsw}.
The spectrum and spatial distribution of this Galactic center excess (GCE) is consistent with a signal expected from DM annihilation,
and consistent with the excess observed in antiprotons.
The second goal which we will pursue in this work is to quantify more precisely the above statement, performing
joint fits of the antiproton and gamma-ray signals for various individual DM annihilation channels.
In performing this comparison we will also use the most recent results of gamma-ray observations from
dwarf satellite galaxies of the Milky Way, which are a known sensitive probe of DM annihilation.

Finally, while DM annihilation can be probed in a rather model-independent way by considering individual SM annihilation channels, it is interesting to also test specific models of DM\@. Such models typically predict CR and gamma-ray fluxes 
from a combination of various SM annihilation channels, and they can be confronted with direct and collider searches for DM\@. As an example, we shall thus consider a minimal Higgs portal  DM model, which adds a real singlet scalar DM field $S$ to the SM\@. 
We shall demonstrate that the scalar Higgs portal model can accommodate both the CR antiproton flux and the GCE, despite strong constraints from invisible Higgs decays and direct DM detection.

The paper is organized as follows. In section~\ref{sec:indchannels} we analyze the CR antiproton data for individual DM annihilation channels. The joint analysis of antiproton and gamma-ray fluxes, including both the GCE and dwarf galaxies, is presented in section~\ref{sec:jointfits}. In section~\ref{sec:HP} we consider the specific case of the scalar Higgs portal model, and present a global analysis 
including antiproton and gamma-ray fluxes, as well as constraints  from the DM relic density, invisible Higgs decays and direct DM searches. We conclude in section~\ref{sec:summary}.

%===================================================================
\section{Cosmic-ray fits for individual dark matter annihilation channels}\label{sec:indchannels}
%===================================================================

DM annihilation in the Galaxy results in a flux of antiprotons from the hadronization and decay of SM particles. The corresponding source term is given by
\begin{eqnarray}
  \label{eqn::DM_source_term}
  q_{\bar{p}}^{(\mathrm{DM})}(\bm{x}, E_\mathrm{kin}) = 
  \frac{1}{2} \left( \frac{\rho(\bm{x})}{m_\mathrm{DM}}\right)^2  \sum_f \left\langle \sigma v \right\rangle_f \frac{\diff N^f_{\bar{p}}}{\diff E_\mathrm{kin}} ,
\end{eqnarray}
where $m_\mathrm{DM}$ is  the  DM mass and $\rho(\bm{x})$ the DM density distribution. 
The  thermally averaged annihilation cross section for the SM final state $f$,  
${\rm DM}\!+\!{\rm DM} \to f\!+\!\bar{f}$, is denoted by $\left\langle \sigma v \right\rangle_f$, and $\diff N^f_{\bar{p}}/\diff E_\mathrm{kin}$ is the corresponding antiproton 
energy spectrum per DM annihilation. Note that the factor $1/2$ corresponds to scalar or Majorana fermion DM.

We use the NFW DM density profile~\cite{Navarro:1995iw}, 
 $   \rho_{\mathrm{NFW}}(r) = \rho_h \, r_h/r\, \left( 1 + r/r_h \right)^{-2}$, 
with a characteristic halo radius $r_h=20\,$kpc, and a characteristic halo density $\rho_h$, normalized to a 
local DM density $\rho_\odot = 0.43\,$GeV/cm$^3$~\cite{Salucci:2010qr} at the 
solar position $r_ \odot = 8\,$kpc. The choice of the DM profile has a negligible impact on our results, 
as demonstrated in \cite{Cuoco:2016eej}.

The energy distribution and yield of antiprotons per DM annihilation, ${\diff N^f_{\bar{p}}}/{\diff E_\mathrm{kin}}$, is determined by the DM mass and  the relevant SM annihilation channel. We use 
the results presented in \cite{Cirelli:2010xx} for the annihilation into gluons, $b \bar b$, $t\bar t$ and $hh$. (The spectra for annihilation into light quarks are very similar to those for gluons.) For $ZZ^{*}$ and $WW^{*}$ final states we have 
generated the spectra with \textsc{Madgraph5\_aMC@NLO}~\cite{Alwall:2014hca} and
\textsc{Pythia~8.215}~\cite{Sjostrand:2007gs}, adopting the default setting and scale choice, $Q=m_\text{DM}$.
We note that the choice of the \textsc{Pythia} tune may introduce uncertainties up to about 15\%,
while varying the shower scale in a range between $m_\text{DM}/6$ and $2m_\text{DM}$ can result in uncertainties of up to 30\%.\footnote{%
Note that in \textsc{Madgraph5\_aMC@NLO} for the default setting (dynamical
scale) the scale is set to $m_\text{DM}/6$.}
This difference is induced through the strength of the final state radiation.
However, we have checked that the theoretical uncertainty in the prediction of the antiproton energy spectrum 
from this scale variation and different \textsc{Pythia} tunes does not affect our results. 
For the default \textsc{Pythia} settings, annihilation spectra into (on-shell) $WW$ and $ZZ$ 
are in reasonable agreement with those of~\cite{Cirelli:2010xx}.

To analyze the impact of DM annihilation on the CR antiproton flux, we perform a joint analysis of the fluxes of protons, helium
and antiprotons, including a potential contribution from DM annihilation, which would affect the antiproton to proton ratio. We solve the standard diffusion equation using \textsc{Galprop} \cite{Strong:1998fr,Strong:2015zva}, 
assuming a cylindrical symmetry for our Galaxy, with a radial extension of $20\,$kpc. 
In total, we analyze a parameter space with thirteen dimensions. Eleven parameters
are related to the CR sources and the propagation of CRs, while 
for each individual SM annihilation channel,  the  DM component of the CR flux is specified  by the DM mass and its annihilation cross section.
The parameters describing the CR sources and propagation, as well as the DM contribution, are determined in a global fit of the AMS-02 proton and helium fluxes \cite{Aguilar_AMS_Proton_2015, Aguilar_AMS_Helium_2015}, and the AMS-02 antiproton to proton ratio \cite{Aguilar:2016kjl}, complemented 
by proton and helium data from CREAM~\cite{Yoon_CREAM_CR_ProtonHelium_2011} and VOYAGER~\cite{Stone_VOYAGER_CR_LIS_FLUX_2013}. We use \textsc{MultiNest} \cite{Feroz_MultiNest_2008} to scan this parameter space and derive the corresponding profile likelihoods. For details of the propagation model and the numerical analysis we refer to~\cite{Korsmeier:2016kha,Cuoco:2016eej}. 

We use as benchmark antiproton production cross section the default in \textsc{Galprop}, i.e., 
the parameterization from~\cite{TanNg_AntiprotonParametrization_1983}.
In~\cite{Cuoco:2016eej} we checked recent new updated models of the cross section 
from~\cite{Mauro_Antiproton_Cross_Section_2014} and~\cite{Kachelriess:2015wpa},
and we found that the results of the fit are substantially unchanged. The main effect is to slightly modify the region of parameter space
preferred by DM at the level of 20--30\%, leaving unchanged the values of the minimal $\chi^2$.

Adding a DM component significantly improves the global fit of the CR antiproton data. This is due to a sharp spectral feature in the antiproton flux at a rigidity of about 20\,GV. Such a feature cannot be described by the smooth spectrum of secondary antiprotons produced by the interactions of primary protons
and helium nuclei on the interstellar medium. The spectrum from DM annihilation, on the other hand, exhibits such a sharp feature from the kinematic cut-off set by the DM mass. Adding a DM component thus provides a significantly better description of the antiproton data. 

%=====================
%    \                                           |
%      \                                         |
%        \                                       |
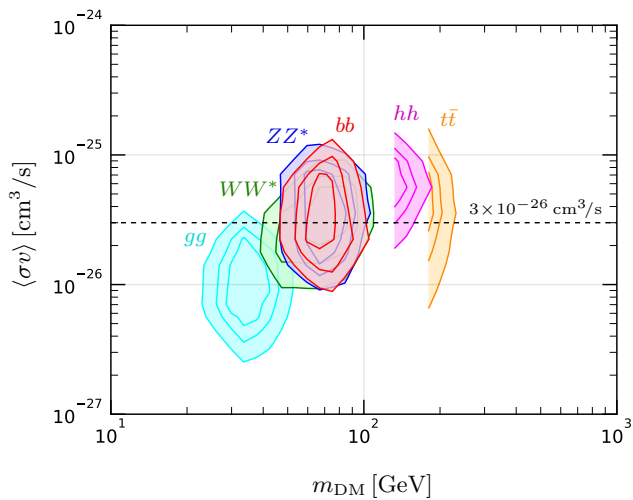
\begin{figure}[!h]
\centering
\setlength{\unitlength}{1\textwidth}
\begin{picture}(0.57,0.43)
 \put(0.0,0.0){ 
  \put(0.0,0.0){\includegraphics[width=0.58\textwidth]{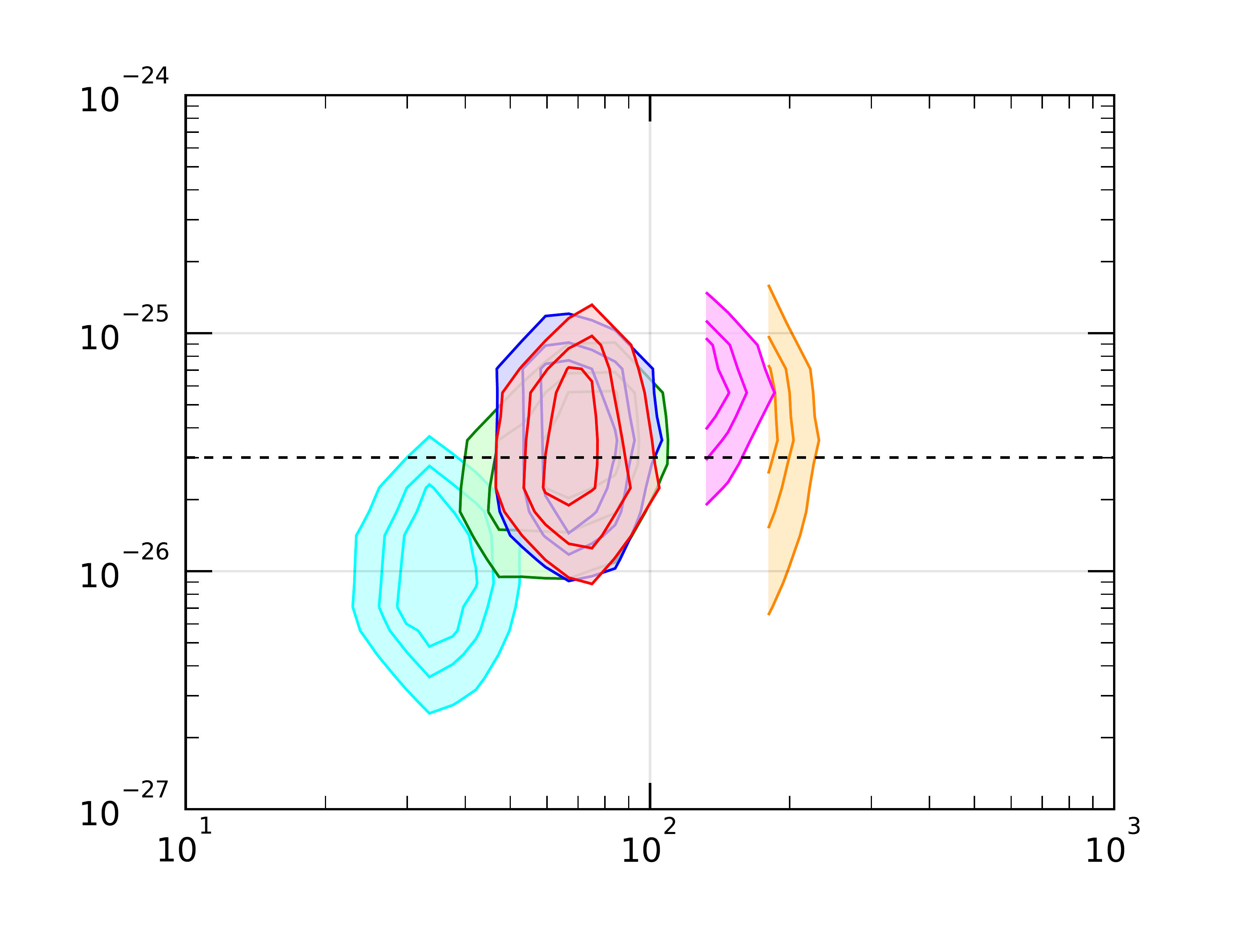}}
  \put(0.26,0.0){\footnotesize $m_\text{DM} \,[\text{GeV}]$}
  \put(0.0,0.175){\rotatebox{90}{\footnotesize $\sv \,[\text{cm}^3/\text{s}]$}}
  \put(0.37,0.315){\scriptsize {\color{plotorange} $t\bar t$}}
  \put(0.33,0.32){\scriptsize {\color{plotpink} $hh$}}
  \put(0.22,0.3){\scriptsize {\color{plotblue} $ZZ^*$}}
  \put(0.28,0.31){\scriptsize {\color{plotred} $bb$}}
  \put(0.18,0.255){\scriptsize {\color{plotgreen} $WW^*$}}
  \put(0.15,0.215){\scriptsize {\color{plotcyan} $gg$}}
  \put(0.395,0.239){\tiny $3\!\times\!10^{-26}\,\text{cm}^{3}\!/\text{s}$}
  }
\end{picture}
\caption{Cosmic-ray fit for the individual annihilation channels: $gg$ (cyan), $WW^*$ (green), 
$b\bar b$ (red), $ZZ^*$ (blue), $hh$ (pink) and $t\bar t$ (orange) in the $m_\text{DM}$-$\sv$ plane.
We show the 1, 2, and 3\,$\sigma$ contours. For comparison we display the thermal cross section (dashed
horizontal line).
}
\label{fig:CRind}
\end{figure}
%                                      \         |
%                                        \       |
%                                          \     |
%=====================

In figure~\ref{fig:CRind} we present the preferred range of DM masses and annihilation cross sections for the different SM annihilation channels. 
The regions are frequentist contour plots of the two-dimensional profile likelihood obtained minimizing the $\chi^2$ with
respect to the remaining eleven parameters in the fit. They, thus, include the uncertainties in the CR source spectra and CR propagation.
All channels provide an improvement compared to a fit without DM: we find a $\chi^2$/(number of degrees of freedom) of 71/165 for the fit without DM, which is reduced to 
46/163 ($b \bar b$), 48/163 ($hh$), 50/163 (gluons and/or light quarks),  50/163 ($WW^*$), 46/163 ($ZZ^*$),  and 59/163 ($t \bar t$), respectively, when adding a corresponding DM component (see also Table~\ref{tab:jointfits}). Formally, $\Delta \chi^2 = 25$ for
the two extra parameters introduced by the DM component with annihilation into $b\bar b$ corresponds to a significance of 4.5, although such an estimate does not account for possible systematic errors. 

Figure~\ref{fig:CRind} also shows that different annihilation channels would imply different preferred DM masses, ranging from $m_{\rm DM} \approx 35$\,GeV for gluons and/or light quarks to $m_{\rm DM}$ near the Higgs and top mass for annihilation into Higgs or top-quark pairs, respectively. For all the channels, the fit points to a thermal annihilation cross section $\left\langle \sigma v \right\rangle \approx 3 \times 10^{-26}$~cm$^3$/s.

It can be noted that the values of the $\chi^2$ are typically quite low for both the fits with and without DM.
This is due to the fact that CR data error-bars are dominated by systematic errors rather than statistical errors.
This is true in particular for the proton and helium data, while for  the antiproton to proton ratio the two errors
have comparable weight.
As a consequence, the use of $\chi^2$ statistics to describe the data is not fully correct.
A proper treatment would require a deeper knowledge of the
systematic uncertainties so that to include them directly at the level of the likelihood rather
than in the error-bars. This information is, however, not publicly available, and
this approach is not possible at the moment.
It would be desirable that such a more complete information
is released  for future CR data publications and updates.

%===================================================================
\section{Joint fit of antiproton and gamma-ray fluxes}\label{sec:jointfits}
%===================================================================

DM annihilation would also result in a flux of gamma rays, predominantly from the decay of pions produced in 
the fragmentation of SM particles. The gamma-ray flux per unit solid angle at a photon energy $E_\gamma$ is 
\begin{equation}\label{eq:gammaflux}
 \frac{\textrm{d}\Phi}{\textrm{d}\Omega\textrm{d}E}   = \frac{1}{2 m_{\rm DM}^2} \sum_{f} \frac{\textrm{d}N^{f}_{\gamma}}{\textrm{d}E} \frac{\langle \sigma v\rangle_{f}}{4 \pi} \int\limits_{\textrm{l.o.s.}}\textrm{d}s\, \rho^2\left(r(s,\theta)\right)\,,
\end{equation}
where  $\textrm{d}{N}^{f}_{\gamma}/\textrm{d}E$ is the photon spectrum per annihilation for a given final  state $f$, and $\langle \sigma v\rangle_{f}$ is the corresponding  annihilation cross section. 
The integral has to be evaluated  along the line-of-sight (l.o.s.) at an observational angle $\theta$ towards the Galactic center. The l.o.s.\ integral of the DM 
density-squared, $\rho^2$, over the solid angle $\textrm{d}\Omega$ is called the $J$-factor. 
We adopt a generalized NFW profile~\cite{Calore:2014xka} with an inner slope $\gamma\simeq1.2$  for the DM density $\rho$. 
This is in contrast to the standard NFW profile applied in section~\ref{sec:indchannels}. 
However, CRs and gamma-rays probe very different parts of the profiles. 
The l.o.s. integral in Eq.~(\ref{eq:gammaflux}) is very sensitive to the profile behavior close to the Galactic center, 
while CRs mostly probe the local DM distribution. 
Indeed, in the latter case we verified that even changing to the cored Burkert profile \cite{Burkert:1995yz} does not affect the results of the CR fit \cite{Cuoco:2016eej}. 
From this point of view it is legitimate to use an NFW profile for CRs while adopting the generalized NFW profile for gamma-rays.

An excess in the flux of gamma rays from the Galactic center has been reported by several groups~\cite{Vitale:2009hr,Goodenough:2009gk,Hooper:2010mq,Hooper:2011ti,Abazajian:2012pn,Hooper:2013rwa,Gordon:2013vta,Abazajian:2014fta,Daylan:2014rsa,Calore:2014xka,TheFermi-LAT:2015kwa,Karwin:2016tsw} (but see also \cite{TheFermi-LAT:2017vmf}).
The GCE is peaked at photon energies of a few GeV, and consistent with a spherical morphology,  extending up to at least~10$^\circ$ away from the Galactic center,   
and a steep radial profile~\cite{Daylan:2014rsa,Calore:2014xka}. 
Various astrophysical processes have been proposed to explain the excess~\cite{Petrovic:2014uda,Petrovic:2014xra,Cholis:2015dea}.
Also, studies based on photon-count statistic suggest that the excess is  more compatible with a population of
unresolved point sources rather than with a pure diffuse emission \cite{Bartels:2015aea,Lee:2015fea,Fermi-LAT:2017yoi}.
Nonetheless, a DM interpretation  is still viable. In particular, the excess is
compatible with the signal expected from the annihilation of DM, with a cross section close to the thermal  value and with a DM mass around 50~GeV.
In our analysis of the GCE we will use the gamma-ray energy spectrum and error covariance matrix obtained in~\cite{Calore:2014xka}.

%=====================
%    \                                           |
%      \                                         |
%        \                                       |
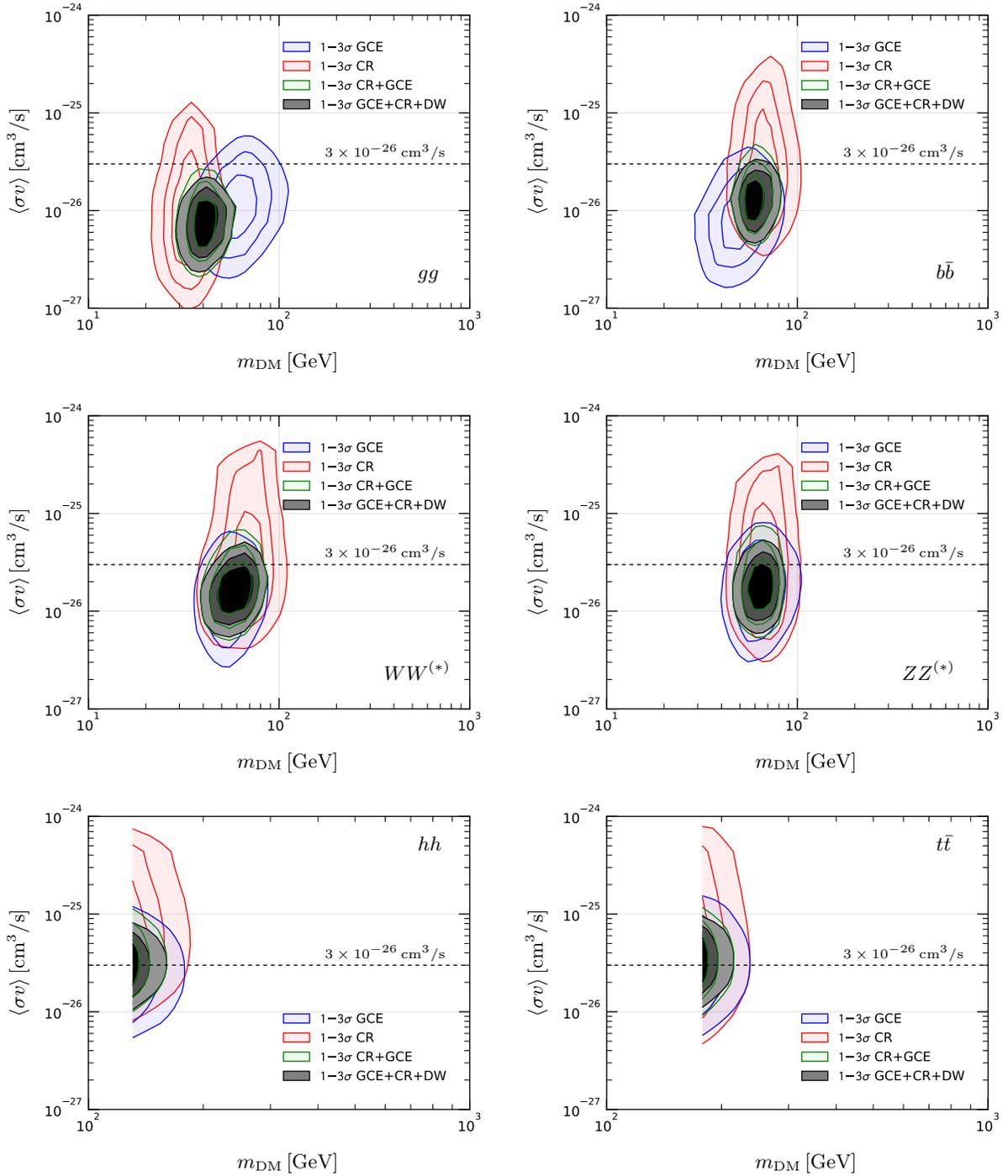
\begin{figure}[h]
\vspace{0.5cm}
\centering
\setlength{\unitlength}{1\textwidth}
\begin{picture}(0.99,1.2)
 \put(-0.01,0.82){ % top left
  \put(0.01,-0.0){\includegraphics[width=0.52\textwidth]{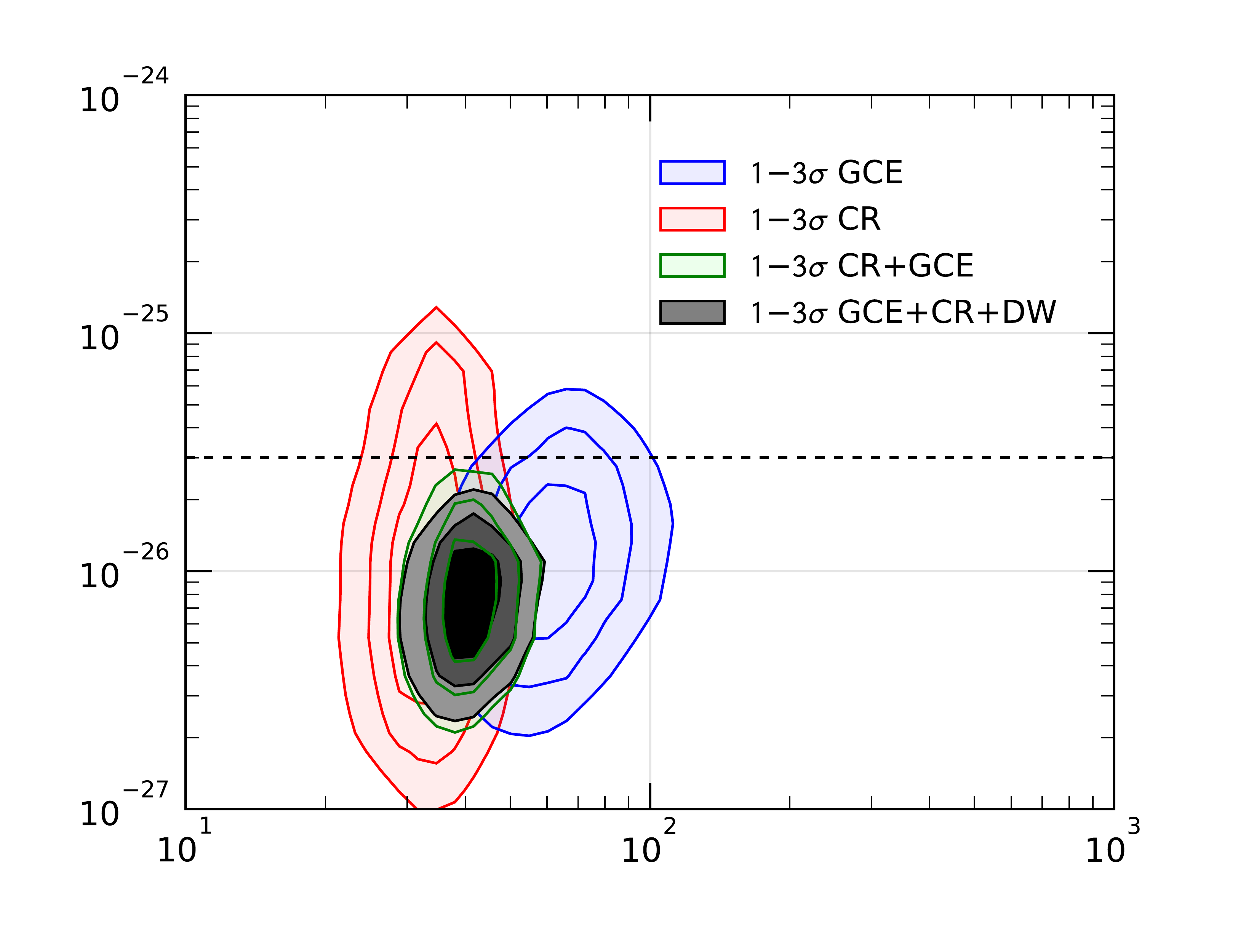}}
  \put(0.24,0.0){\footnotesize $m_\text{DM} \,[\text{GeV}]$}
  \put(0.33,0.215){\tiny $3\times10^{-26}\,\text{cm}^{3}/\text{s}$}
  \put(0.005,0.155){\rotatebox{90}{\footnotesize $\sv \,[\text{cm}^3/\text{s}]$}}
  \put(0.425,0.09){\footnotesize $gg$}
  }
 \put(0.52,0.82){ % top right
  \put(0.01,-0.0){\includegraphics[width=0.52\textwidth]{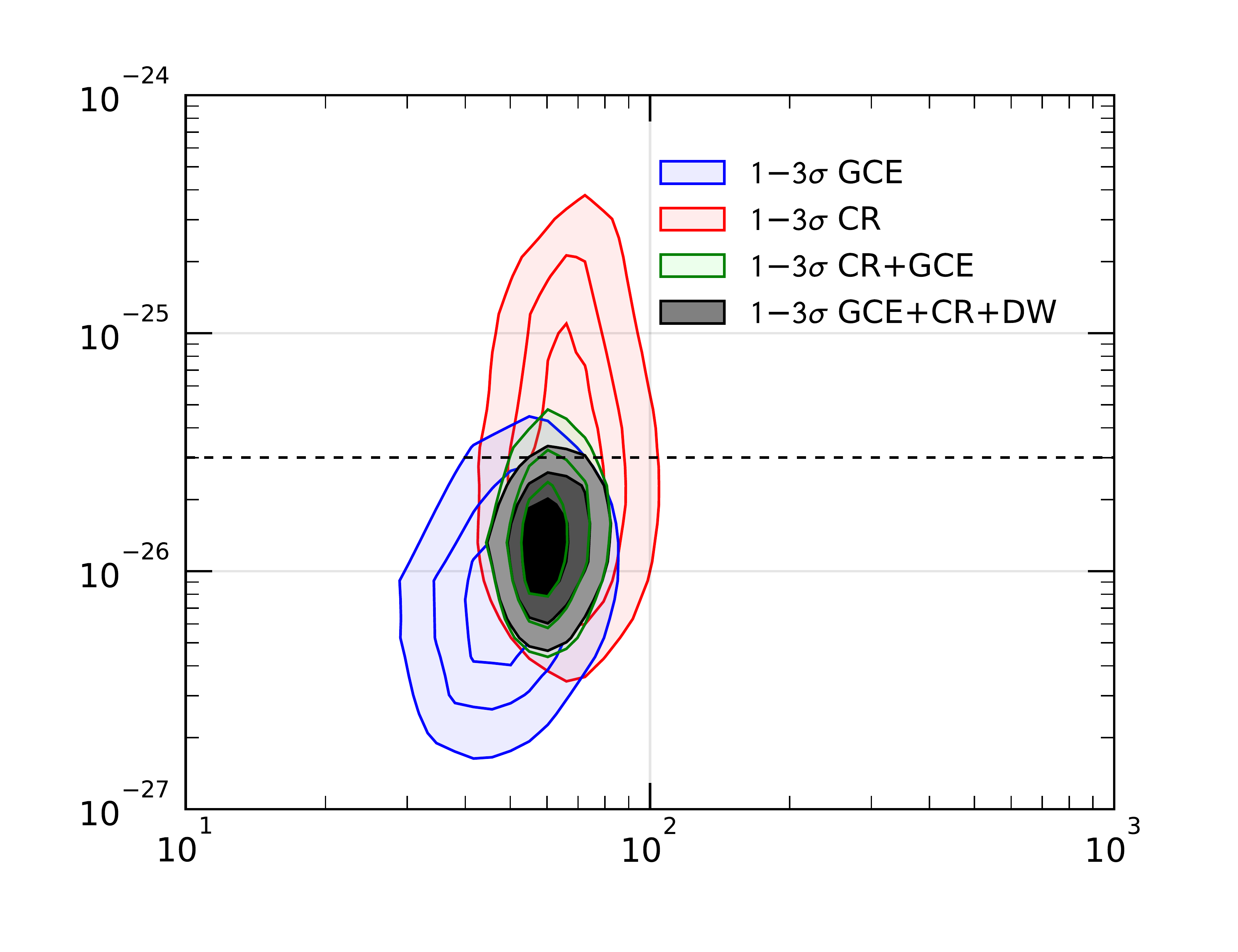}}
  \put(0.24,0.0){\footnotesize $m_\text{DM} \,[\text{GeV}]$}
  \put(0.33,0.215){\tiny $3\times10^{-26}\,\text{cm}^{3}/\text{s}$}
  \put(0.005,0.155){\rotatebox{90}{\footnotesize $\sv \,[\text{cm}^3/\text{s}]$}}
  \put(0.425,0.09){\footnotesize $b\bar b$}
  }
 \put(-0.01,0.41){ % middle left
  \put(0.01,-0.0){\includegraphics[width=0.52\textwidth]{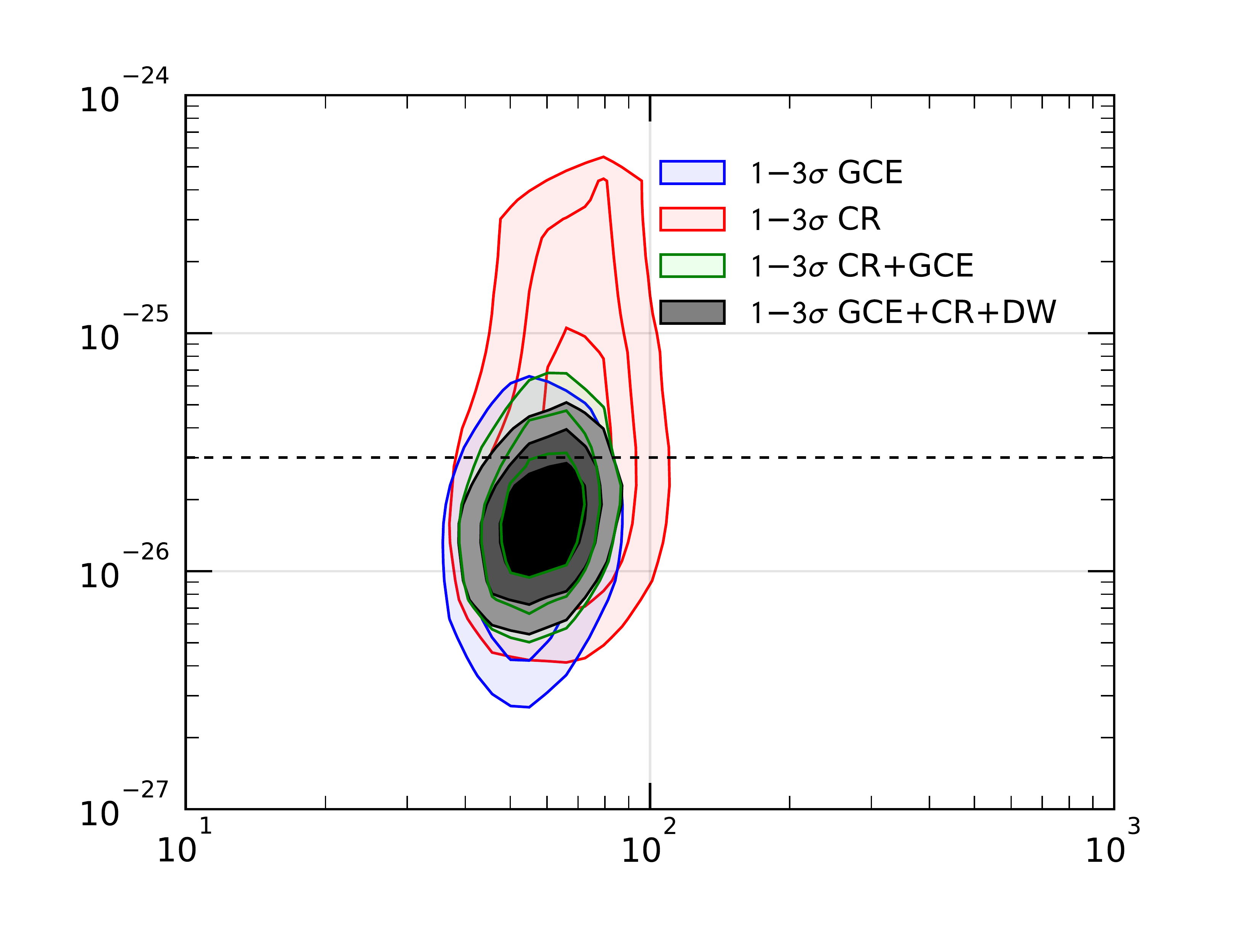}}
  \put(0.33,0.215){\tiny $3\times10^{-26}\,\text{cm}^{3}/\text{s}$}
  \put(0.24,0.0){\footnotesize $m_\text{DM} \,[\text{GeV}]$}
  \put(0.005,0.155){\rotatebox{90}{\footnotesize $\sv \,[\text{cm}^3/\text{s}]$}}
  \put(0.39,0.09){\footnotesize $WW^{(*)}$}
  }
 \put(0.52,0.41){ % middle right
  \put(0.01,-0.0){\includegraphics[width=0.52\textwidth]{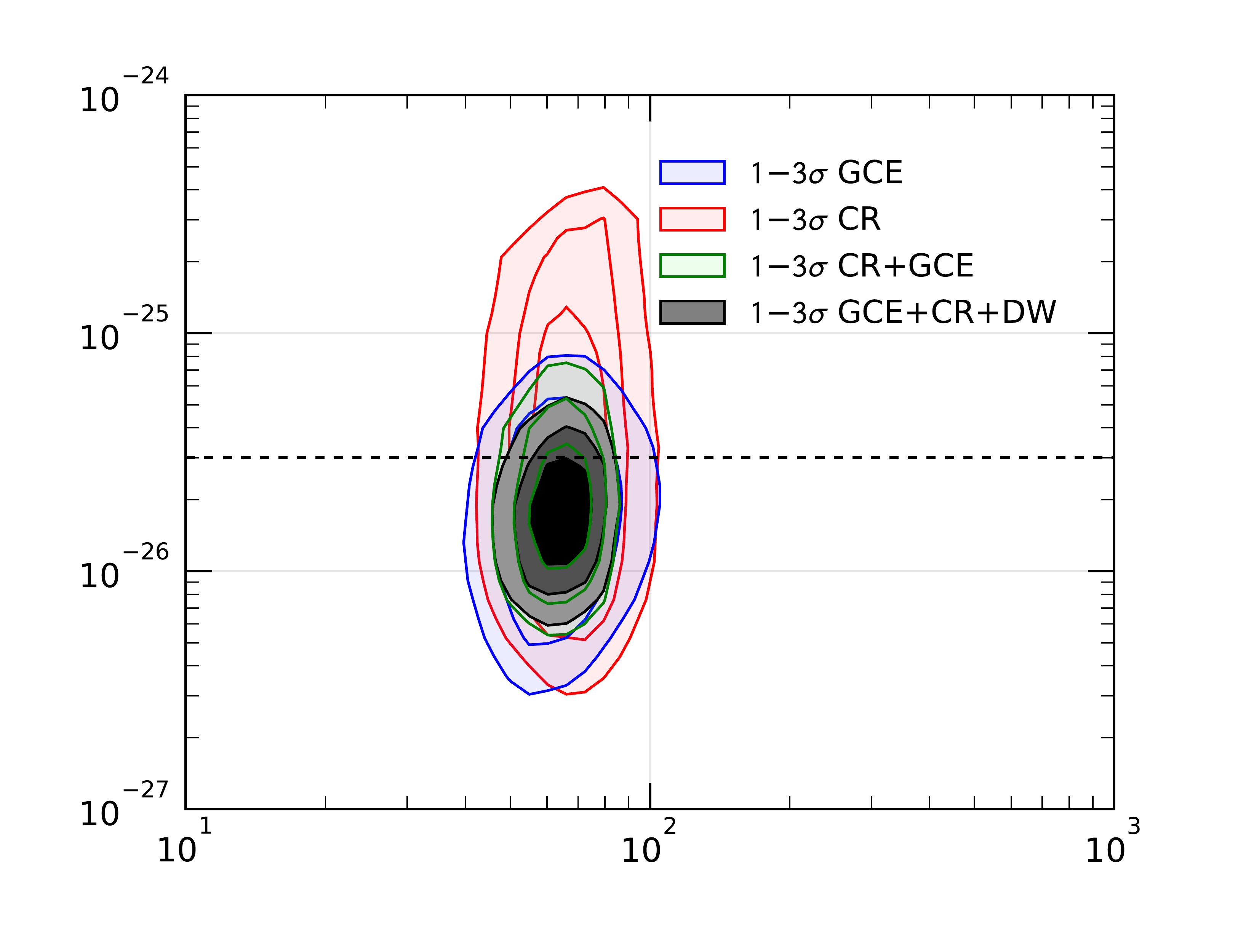}}
  \put(0.33,0.215){\tiny $3\times10^{-26}\,\text{cm}^{3}/\text{s}$}
  \put(0.24,0.0){\footnotesize $m_\text{DM} \,[\text{GeV}]$}
  \put(0.005,0.155){\rotatebox{90}{\footnotesize $\sv \,[\text{cm}^3/\text{s}]$}}
  \put(0.39,0.09){\footnotesize$ZZ^{(*)}$}
  }
 \put(-0.01,0.0){ % bottom left
  \put(0.01,-0.0){\includegraphics[width=0.52\textwidth]{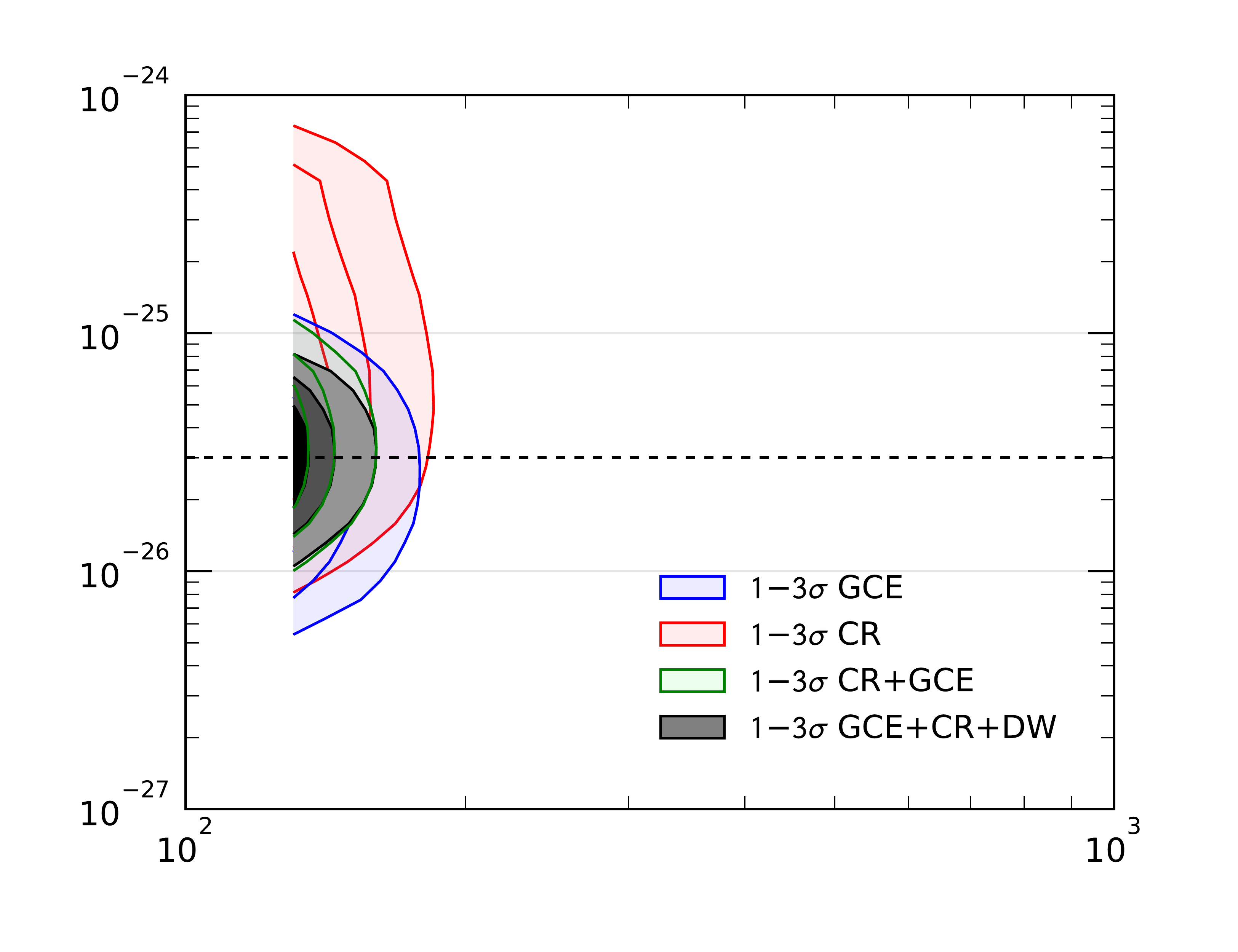}}
  \put(0.33,0.215){\tiny $3\times10^{-26}\,\text{cm}^{3}/\text{s}$}
  \put(0.24,0.0){\footnotesize $m_\text{DM} \,[\text{GeV}]$}
  \put(0.005,0.155){\rotatebox{90}{\footnotesize $\sv \,[\text{cm}^3/\text{s}]$}}
  \put(0.425,0.325){\footnotesize $hh$}
  }
 \put(0.52,0.0){ % bottom right
  \put(0.01,-0.0){\includegraphics[width=0.52\textwidth]{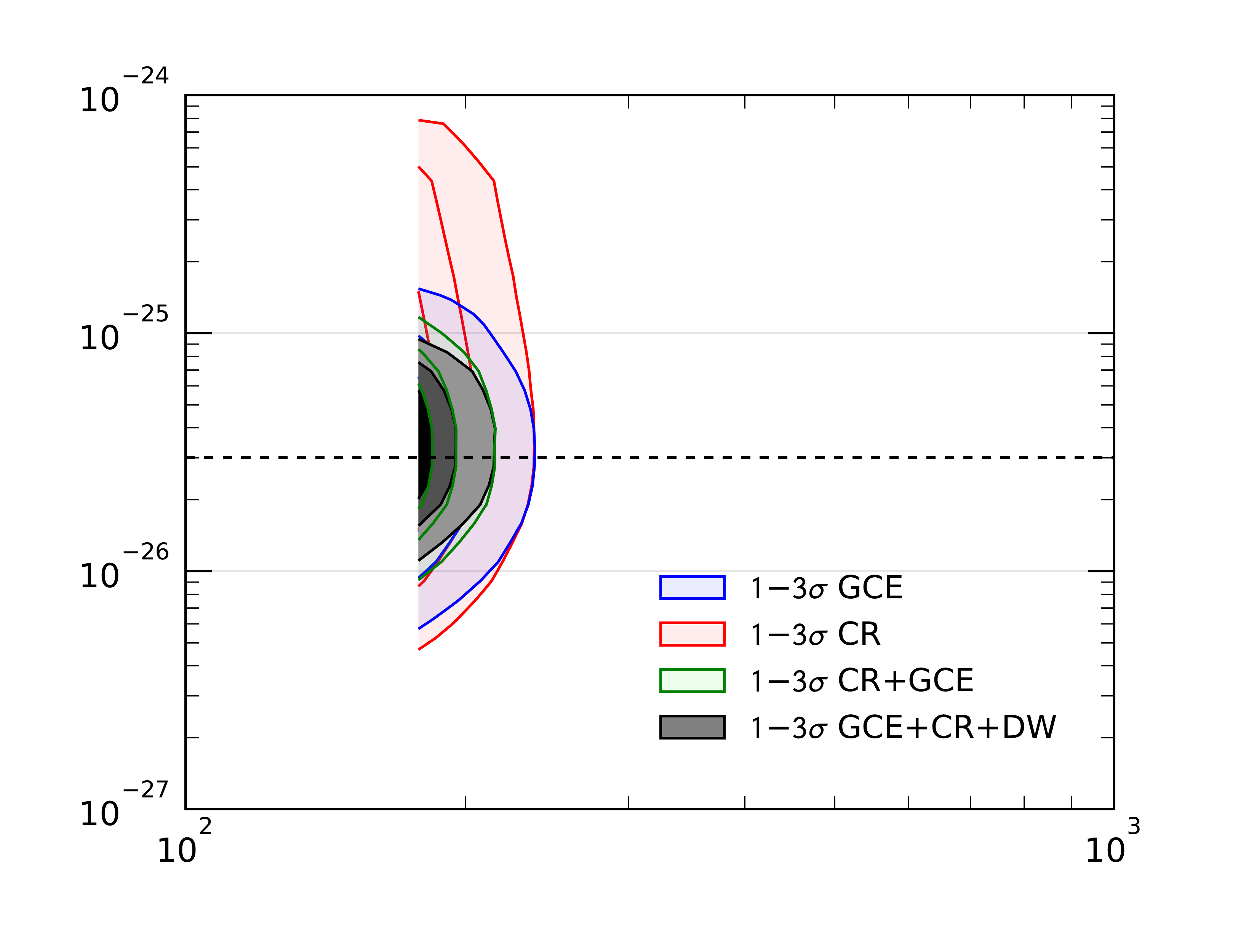}}
  \put(0.33,0.215){\tiny $3\times10^{-26}\,\text{cm}^{3}/\text{s}$}
  \put(0.24,0.0){\footnotesize $m_\text{DM} \,[\text{GeV}]$}
  \put(0.005,0.155){\rotatebox{90}{\footnotesize $\sv \,[\text{cm}^3/\text{s}]$}}
  \put(0.425,0.325){\footnotesize $t\bar t$}
  }
\end{picture}
\vspace{0.5cm}
\caption{Joint fit to CR fluxes, the GCE and dwarf galaxies for the individual SM annihilation channels in the $m_\text{DM}$-$\sv$ plane.
We show the 1, 2, and 3\,$\sigma$ contours. For comparison we display the thermal cross section (dashed
horizontal line).
}
\label{fig:jointind}
\end{figure}
%                                      \         |
%                                        \       |
%                                          \     |
%=====================

DM annihilation in gamma rays can also be sensitively tested by observations of  dwarf satellite galaxies of the Milky Way 
\cite{Ackermann:2013yva,Drlica-Wagner:2015xua,Ackermann:2015zua,Fermi-LAT:2016uux}.
Here, we use the likelihood as a function of the flux for each dwarf provided by 
Fermi-LAT~\cite{Fermi-LAT:2016uux}, and the gamma-ray spectra for the individual annihilation channels obtained in~\cite{Cuoco:2016jqt}.
We consider a total of eleven dwarfs: the seven brightest confirmed dwarfs analyzed in~\cite{Fermi-LAT:2016uux} 
(Coma Berenices, Draco, Sculptor, Segue 1, Ursa Major II, Ursa Minor, Reticulum II) as well as 
Willman 1, Tucana III, Tucana IV and Indus II. Four of these dwarfs (Reticulum II, Tucana III, Tucana IV, Indus II)
exhibit small excesses at the level of $\sim2\sigma$ (local) each, which are compatible with a signal from DM
annihilation with a thermal cross section~\cite{Geringer-Sameth:2015lua,Li:2015kag,Fermi-LAT:2016uux}. The total likelihood is obtained as a product of likelihoods over each single dwarf as described in \cite{Ackermann:2015zua,Ahnen:2016qkx}. 
The likelihood of each dwarf contains a factor from the flux likelihood,
and a log-normal factor from a deviation of the $J$-factor from its nominal value. 
For the seven brightest confirmed dwarfs we use the $J$-factors and corresponding uncertainties provided in \cite{Fermi-LAT:2016uux} 
(which, in turn, draws  from \cite{Geringer-Sameth:2014yza}, except for Reticulum II, whose $J$-factor is taken from \cite{Simon:2015fdw}),
while for Willman we use the $J$-factor from \cite{Ackermann:2015zua}. For 
Tucana III, Tucana IV and Indus II we use the distance-based predictions provided 
in~\cite{Fermi-LAT:2016uux} with a medium estimated error of 0.6 dex.
We marginalize over the $J$-factors of the individual dwarf galaxies in the fit.
It should be noted that estimates of the $J$-factors present some differences depending on the analysis 
(compare for example \cite{Geringer-Sameth:2014yza,Martinez:2013els}), with some analysis \cite{Bonnivard:2015xpq}
 finding somewhat lower values than the others.
This, however, has only  a minor impact on our results, since, as we show below, 
the results of the fits are dominated by the GCE and CR signals.

On the basis of the likelihoods obtained in the CR fit described in section~\ref{sec:indchannels}
we now perform a joint fit of CR antiprotons and of gamma-rays from
the Galactic center and from dwarf galaxies. 
The gamma-ray fit follows the methodology described in \cite{Cuoco:2016jqt}. 
The fit contains four input parameters, the model parameters, 
$\langle \sigma v \rangle$ and $m_\text{DM}$, as well as the $J$-factor for the Galactic center,
$\log \J$, and the local DM density $\rho_\odot$. 
The latter two parameters are, in principle, not independent.
However, as already mentioned above, CRs and gamma-rays probes different
parts of the DM distribution in the Galaxy and it is thus reasonable to
explore  the uncertainties in these two parameters as independent.
For  $\log \J$  we use a gaussian distribution (log-normal in $\J$)
with mean 53.54 and error 0.43, i.e., $\log (J/\text{GeV}^{2}\text{cm}^{-5})=53.54\pm 0.43$.
This GC $J$-factor  refers to an integration region of $40^\circ \times 40^\circ$ around the GC
and with a stripe of $\pm 2^\circ$ masked along the Galactic plane,
in order to be compatible with the GCE data from Ref.~\cite{Calore:2014xka} that we use.
The details of the derivation of  the distribution in $\log \J$ and the error  
are described in~\cite{Cuoco:2016jqt}.
For the local DM density we also use Gaussian errors
$\rho_\odot=0.43\pm0.15$ GeV/cm$^3$~\cite{Salucci:2010qr}.
We use the result of Ref.~\cite{Salucci:2010qr} in order to be conservative since $\rho_\odot$ has a 
relatively large error.  A recent review on the status of the determination of $\rho_\odot$
is given in~\cite{Read:2014qva}.

 Figure~\ref{fig:jointind} shows the preferred range of DM masses and annihilation cross sections, 
where we have marginalized over $\log \J$ and $\rho_\odot$. We present 1, 2, and 3\,$\sigma$ contours for a fit to the GCE (blue), CR (red), CR+GCE (green) and
CR+GCE+dwarfs (black) for the six annihilation channels $gg$, $b \bar b$, $WW^{(*)}$,
$ZZ^{(*)}$, $hh$ and $t\bar t$. Note that the fits to the CR fluxes in figure~\ref{fig:jointind} show a wider spread in $\langle \sigma v \rangle$ than those displayed in figure~\ref{fig:CRind}, because in figure~\ref{fig:jointind} we marginalize over the local DM density, $\rho_\odot=0.43\pm0.15$ GeV/cm$^3$,
while in figure~\ref{fig:CRind} a fixed value $\rho_\odot=0.43$ GeV/cm$^3$ is used.

For most SM annihilation channels, we observe very good agreement between the DM interpretation of the CR antiprotons and the GCE gamma-ray flux. The preferred region in $\langle \sigma v \rangle$ and $m_\text{DM}$ is consistent when comparing the CR and GCE fits individually, and the combined CR+GCE fit. However, as can be seen in the upper left panel of figure~\ref{fig:jointind}, annihilation into gluons (or light quarks) is disfavored as an explanation of both the CR antiproton flux and the GCE, as both signals individually prefer different regions of DM mass. 
Annihilation into $t$ quarks is also disfavored since it does not provide a good fit to either the GCE or antiprotons.
Adding the constraints from dwarf galaxies  disfavors large values for $\langle \sigma v \rangle$, but hardly affects 
the combined CR+GCE fit.  Numerical values of the best-fit $\chi^2$ are reported in Table~\ref{tab:jointfits}.

From the figure we note also that CR prefer a somewhat larger $\sv$ than the GCE and, hence, 
the joint fit pushes $\rho_\odot$ towards slightly larger values with respect to the assumed prior 
from~\cite{Salucci:2010qr}. 
More precisely, we find, with some variation depending on the DM channel, that
the global fit gives a value  $\rho_\odot=0.55\pm0.15$ GeV/cm$^3$, i.e.,
$\sim 0.1$ GeV/cm$^3$ higher than the input prior.
We find that $\rho_\odot=0.3$ GeV/cm$^3$ is at the lower edge of
$\sim 3 \sigma$ range preferred by the fit. 
This means that if the true $\rho_\odot$ is significantly lower than 0.3 GeV/cm$^3$
it becomes difficult to reconcile the GCE with the CR data.
Nonetheless, we also note that for $z_h$, the half-height of the CR propagation region
in the Galaxy, we use the prior 2-7\,kpc, but, since this parameter is unconstrained by the fit (see \cite{Korsmeier:2016kha,Cuoco:2016eej}),  
values up to 10 kpc or more are  allowed.
Thus, since the DM signal approximately scales linearly in $z_h$, higher
$z_h$ values would allow, in consequence, lower $\rho_\odot$ values, making possible a joint fit
of the GCE and CRs down to a  $\rho_\odot$ value of 0.2 GeV/cm$^3$.
This issue is also further discussed in the next section within the Higgs portal fit.

\begin{table}[h]
\begin{center}
\renewcommand{\arraystretch}{1.2}
\begin{tabular}{c | c  c  | c c } 
 & \multicolumn{2}{c |}{individual fits} & \multicolumn{2}{c}{joint fit}\\
 \hline
channel  & 
 $\chi^2_\text{CR}$  &  $\chi^2_\text{GCE}$  & $\chi^2_\text{CR}$  &  $\chi^2_\text{GCE}$  \\ \hline\hline 
 $gg$   &     50.3   &          20.8   &              52.0   &            31.6 \\
 $b\bar b$   &     45.8   &          21.2   &              47.9   &            23.5\\
 $WW^{(*)}$   &     50.4   &          25.6          &       54.6          &     25.6\\
 $ZZ^{(*)}$   &  45.6      &   25.0               &    45.8         &   25.9    \\
$ hh $  &     47.6   &          25.8          &       48.4          &     25.8\\
 $t\bar t$   &     59.5   &          41.1          &       59.5          &     41.1
\end{tabular}
\renewcommand{\arraystretch}{1}
\end{center}
\caption{
$\chi^2$ for the individual fits to CR and GCE as well as for the joint fit.
The number of degrees of freedom for the CR and GCE fit is 163 and 22, respectively.
}
\label{tab:jointfits}
\end{table}

%===================================================================
\section{Interpretation within the singlet scalar Higgs portal model}\label{sec:HP}
%===================================================================

We now discuss a specific minimal model of DM, where we add a singlet scalar field $S$ to the 
SM~\cite{Silveira:1985rk,McDonald:1993ex,Burgess:2000yq}. 
We will follow the analysis in \cite{Cuoco:2016jqt},\footnote{%
An interpretation of the GCE within the singlet Higgs portal model has also been discussed in~\cite{Okada:2013bna,Basak:2014sza,Duerr:2015bea}.}
 with the main difference that now we  include  CR data.

The scalar field interacts with the SM Higgs field $H$ through the Higgs portal operator 
$S^2H^\dagger H$.
Imposing an additional $Z_2$ symmetry, $S \to -S$, the scalar particle is stable and thus a DM candidate. The Lagrangian of the scalar Higgs portal model reads 
\begin{equation}
{\cal L} = {\cal L}_\text{SM} + \frac 12 \partial_\mu  S  \partial^\mu  S  - \frac12 m_{S,0}^2S^2- \frac 14 \lambda_S  S^4- \frac 12 \lambda_{H\!S}\, S^2 H^\dagger H\,.
\label{eq:lagr}
\end{equation}
After electroweak symmetry breaking, the last three terms of the above Lagrangian become
\begin{equation}
{\cal L} \supset  - \frac12 m_{S}^2\, S^2- \frac 14 \lambda_S\,  S^4 - \frac 14 \lambda_{H\!S}\, h^2 S^2 - \frac {1}{2} \lambda_{H\!S}\, v h S^2\,,
\label{eq:ewbr}
\end{equation}
with $H = (h+v, 0)/\sqrt{2}\,$, $v = 246\,$GeV, and where we introduced the physical mass of the singlet 
field, $m_S^2  = m_{S,0}^2 + \lambda_{H\!S} \,v^2 / 2$. The phenomenology of the singlet Higgs portal model has been extensively studied in the literature, see e.g.\ the recent reviews~\cite{Cline:2013gha,Beniwal:2015sdl} and references therein. 

While the scalar self-coupling, $\lambda_S$, is of importance for the stability of the electroweak vacuum, the DM phenomenology of the scalar Higgs portal model is fully specified by the mass of the scalar DM particle, $m_S=m_{\rm DM}$, and the strength of the coupling between the DM and Higgs particles, $\lambda_{H\!S}$. Even though the model is minimal, the $S^2 H^\dagger H$ interaction term implies a rich phenomenology, including invisible Higgs decays, $h \to SS$, a DM-nucleon interaction through the exchange of a Higgs particle, and DM annihilation through 
$s$-channel Higgs, $t$-channel scalar exchange, and the $S^2 h^2$ interactions.

The region most relevant for the DM interpretation of the CR antiproton flux and the GCE is the region 
$m_S\lesssim100\,\text{GeV}$. As this is below the Higgs-pair threshold, $m_S < m_h$,
annihilation proceeds through $s$-channel Higgs exchange only, and 
the relative weight of the different SM final states is determined by the SM Higgs branching ratios, 
independent of the Higgs-scalar coupling $\lambda_{H\!S}$. 
Above the Higgs-pair threshold, $m_S \ge m_h$, the $hh$ final state opens up. The strength of the annihilation into Higgs pairs, as compared to $W,Z$ or top-quark pairs, depends on the size of the Higgs-scalar coupling $\lambda_{H\!S}$. However, as shown in~\cite{Cuoco:2016jqt}, within the scalar Higgs portal model the region above the Higgs-pair threshold that provides a good fit to the GCE (and to the CR) requires very large $\lambda_{H\!S}$ which are  
excluded by direct detection limits. In the following, we will thus focus on DM masses $m_{\rm DM} < m_h$. 

\begin{figure}[h]
\centering
\setlength{\unitlength}{1\textwidth}
\begin{picture}(0.53,0.4)
 \put(-0.01,0.0){ 
  \put(0.01,-0.0){\includegraphics[width=0.52\textwidth]{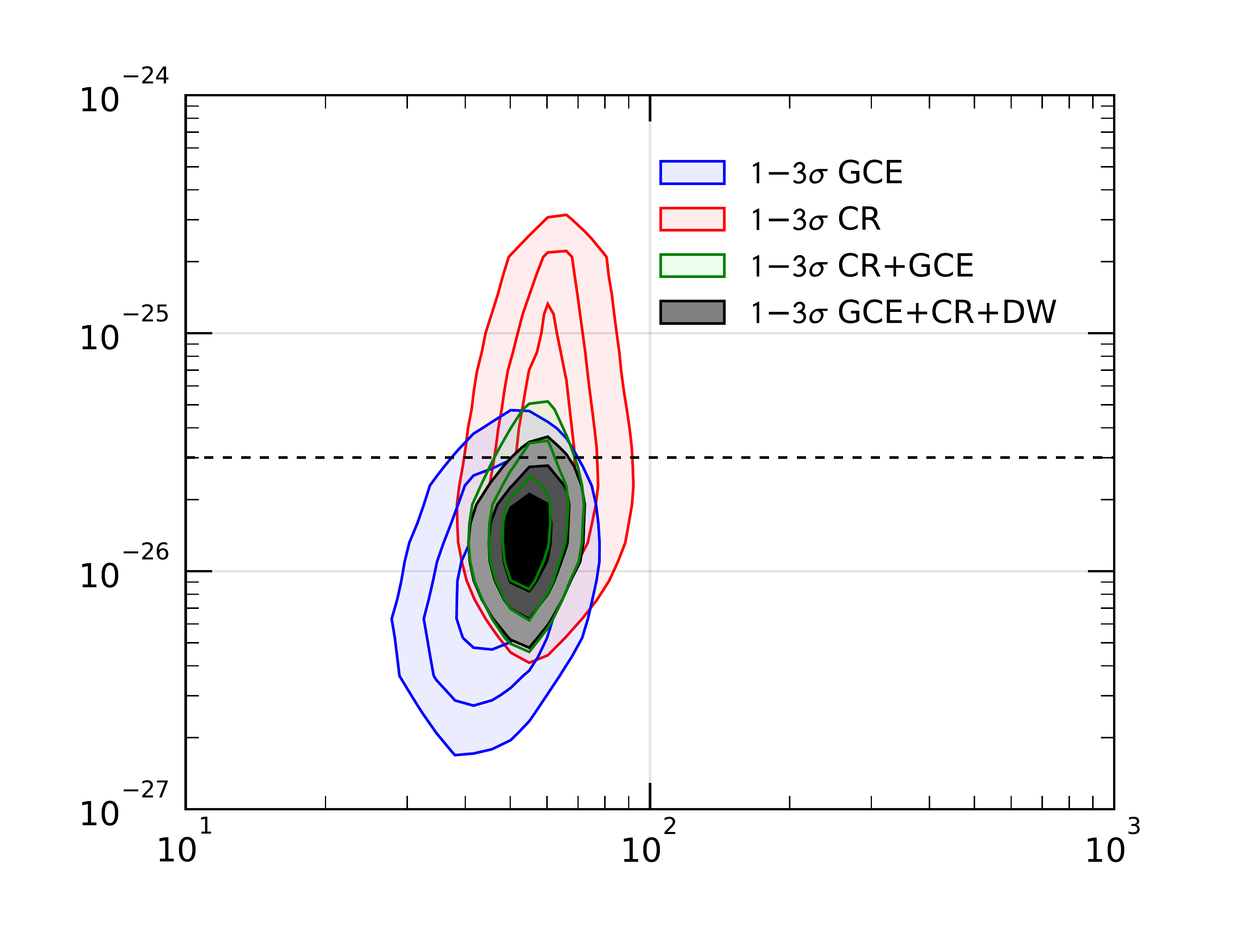}}
  \put(0.33,0.215){\tiny $3\times10^{-26}\,\text{cm}^{3}/\text{s}$}
  \put(0.24,0.0){\footnotesize $m_\text{DM} \,[\text{GeV}]$}
  \put(0.005,0.155){\rotatebox{90}{\footnotesize $\sv \,[\text{cm}^3/\text{s}]$}}
  \put(0.33,0.085){\footnotesize Higgs portal}
  }
\end{picture}
\caption{Joint fit to CR, GCE and dwarfs for the singlet scalar Higgs portal model.
}
\label{fig:jointportal}
\end{figure}
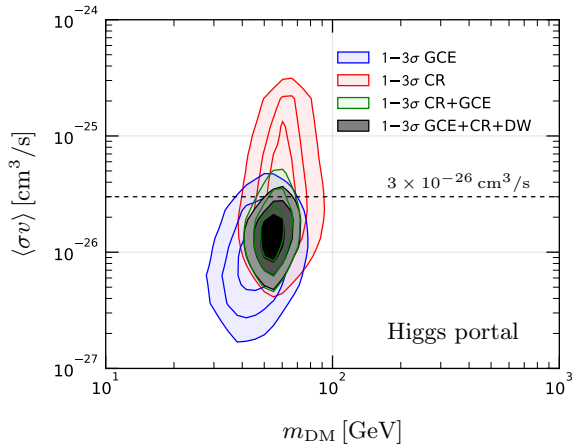

We pursue two approaches. We first adopt a more model-independent point of view and consider a DM interpretation in terms of $m_\text{DM}$ and $\sv$. The only reference 
to the Higgs portal model is through the relative weight of the different SM final states, which is determined by $m_\text{DM}$. Such an analysis probes whether a certain 
combination of annihilation channels, considered individually in section 
\ref{sec:indchannels} and \ref{sec:jointfits}, can provide a fit of the observations. 
Note that this kind of analysis can, in general, not be performed based on the results presented for the individual channels. 
Instead, we perform a dedicated fit to the CR antiproton flux, 
constructing the injection spectra from the spectra of the individual channels according to their relative weights. 
The result is shown in  figure~\ref{fig:jointportal} (red contours), where we have marginalized over  $\rho_\odot$ and $\log \J$. 
The preferred region of DM masses is around $m_{\rm DM} \approx 60$\,GeV, 
where the Higgs portal model predicts annihilation pre-dominantly into bottom quarks, 
$W$-bosons and gluons with a weight of approximately 70, 20 and 10\%, respectively. 
We find a $\chi^2$/(number of degrees of freedom) of 47/163 for the Higgs portal model fit, compared to 71/165 for the fit without DM.

%=====================
%    \                                           |
%      \                                         |
%        \                                       |
%                                      \         |
%                                        \       |
%                                          \     |
%=====================

Performing a joint fit of the CR antiproton flux with the GCE (green contours) as well as with the GCE and dwarf galaxies (black contours)
shifts the preferred region to slightly smaller masses $m_{\rm DM} \approx 55$\,GeV, with a 
$\chi^2$/(number of degrees of freedom) of 49/163 for the CR and $20.8/22$ for the GCE. Although
the best-fit point for the GCE-only fit lies at smaller masses, around $m_{\rm DM} \approx 45$\,GeV (cf.~\cite{Cuoco:2016jqt}), the $\chi^2$/(number of degrees of freedom) for the GCE in the joint fit is almost as good as for the GCE-only fit (which yields $19.2/22$). We can draw the quite general conclusion that DM models where the annihilation is pre-dominantly into $b\bar{b}, WW^{(*)}$ or $ZZ^{(*)}$ final states, or any combination thereof, provide a very good fit of the CR antiproton flux, the GCE and gamma-rays from dwarf galaxies, and point to a DM mass in the vicinity of $m_{\rm DM} \approx 60$\,GeV.

We proceed with a more detailed analysis of the scalar Higgs portal model, where we take into account the various constraints on the parameter space from the Higgs invisible decay width, direct detection searches, searches for gamma-ray lines from the inner Galaxy and the DM relic density. Hence we consider the actual model parameters $m_S=m_{\rm DM}$ and $\lambda_{HS}$ defined in eq.~\eqref{eq:lagr}.

%=====================
%    \                                           |
%      \                                         |
%        \                                       |
\begin{figure}[t]
\centering
\setlength{\unitlength}{1\textwidth}
\begin{picture}(1,1)
 \put(0.0,-0.04){\includegraphics[width=1.0\textwidth]{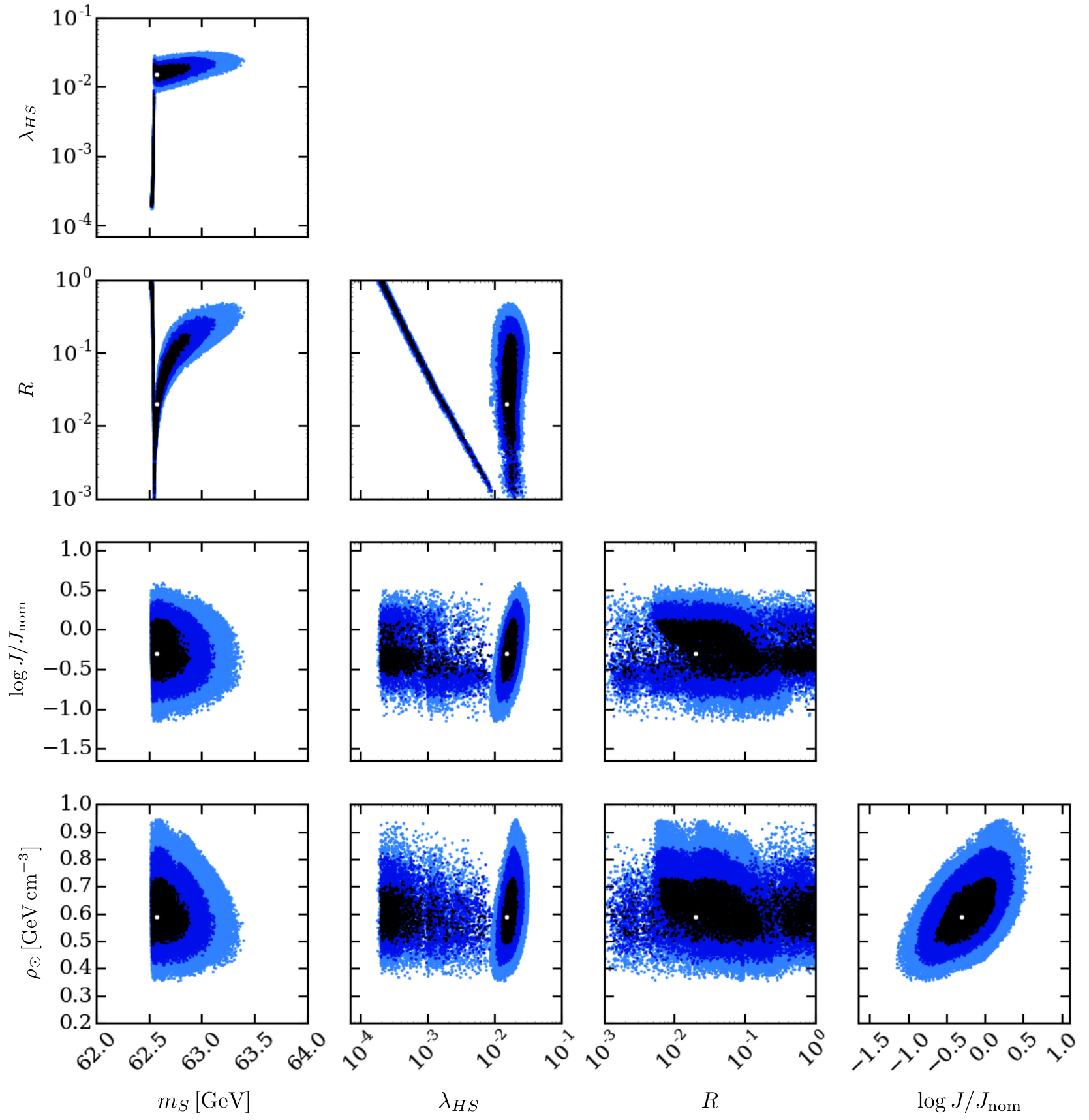}}
\end{picture}
\caption{Triangle plot of the fit to the CR and GCE and all other constraints 
(including unconfirmed dwarfs)
on the parameter space within the Higgs portal singlet scalar model.
The black, blue and light blue points lie within the
1, 2, and $3\sigma$ region around the best-fit point (denoted by a white dot), respectively.
}
\label{fig:HPrepr}
\end{figure}
%                                      \         |
%                                        \       |
%                                          \     |
%=====================

We shall discuss the various constraints briefly in turn, and refer to~\cite{Cuoco:2016jqt} for more details. 
\begin{itemize}
\item For light DM below the Higgs threshold, $m_{\rm DM} < m_h/2$, the invisible Higgs decay $h \to SS$ is kinematically allowed. The LHC limits on the Higgs invisible branching ratio, ${\rm BR}_{\rm inv} \lesssim 0.23$~\cite{Aad:2015pla}, thus  imply an upper limit on the Higgs-scalar coupling $\lambda_{H\!S}$ as a function of the DM mass. 

\item The scalar Higgs portal model predicts a spin-independent DM-nucleon scattering cross section, $\sigma_\text{SI} \propto \lambda_{H\!S}^2/m_{\rm DM}^2$, through the exchange of the SM Higgs boson. The model is therefore severely constrained by direct detection experiments. We use the recent direct detection limits from LUX~\cite{Akerib:2016vxi} in our numerical analysis, updating the results presented in~\cite{Cuoco:2016jqt}. Furthermore,
we introduce the local DM density $\rho_\odot$, relevant for the DM-nucleon scattering rate and the CR flux, as an additional nuisance parameter in the fit.

\item Searches for gamma-ray lines provide constraints on the cross section for the annihilation into mono-chromatic photons, $\langle \sigma v \rangle_{\gamma\gamma}$. We have calculated $\langle \sigma v \rangle_{\gamma\gamma}$ using an Higgs effective Lagrangian as described in~\cite{Cuoco:2016jqt}, and constrain the model with data from the recent Fermi-LAT 
search for spectral lines in the Milky Way halo~\cite{Ackermann:2015lka}. 

\item We require that the Higgs portal model provide the correct DM relic density as measured 
by Planck, $\Omega h^2|_\text{DM} = 0.1198\pm 0.0015$~\cite{Ade:2015xua}. We assume a standard cosmological history, but allow for the possibility that the dark sector is more complex than 
assumed within our minimal model. Hence, the DM density provided by the scalar Higgs portal model is a certain fraction, $R\leq1$, of the density of all gravitationally interacting DM, $\rho_\text{Higgs portal} = R \,\rho_{\text{DM}}$. The total DM density predicted by our model is then $\Omega h^2|_\text{DM} = \Omega h^2|_\text{Higgs portal}\,/R$. We will consider $R$ as a free parameter in our  fit. Note that the annihilation signal today scales as $\propto R^2$, while the direct detection limits scale $\propto R$, thus implying a non-trivial interplay of the various constraints for $R\neq 1$. 
\end{itemize}

In figure~\ref{fig:HPrepr} we present a fit of the Higgs portal model to the CR antiproton flux and the GCE, including the constraints from dwarf galaxies and searches for gamma-ray lines, the invisible Higgs branching ratio, direct DM detection, and the relic density.\footnote{Compared to the analysis presented in \cite{Cuoco:2016jqt} we have included the likelihood of the CR antiproton flux and updated the direct detection and dwarf galaxy limits.}

Let us first consider the upper left panel, which shows the allowed region in the Higgs portal coupling, $\lambda_{HS}$, and the DM mass. The overall flux of antiprotons and photons scales with the annihilation cross section $\langle \sigma v \rangle \propto \lambda_{HS}^2/[(m_h^2-4 m_{\rm DM}^2)^2+\Gamma_h^2 m_h^2]$, where $\Gamma_h$ is the Higgs width. To accommodate the CR data and the GCE, either large couplings $\lambda_{HS}$ or masses near the Higgs resonance, $m_{\rm DM}\approx m_h/2$, are required. However, large couplings are excluded by the invisible Higgs branching ratio for masses $m_{\rm DM} \lesssim m_h/2$, and by direct detection limits for masses $m_{\rm DM} \gtrsim m_h/2$, leaving only the region near the Higgs threshold $m_{\rm DM}\approx m_h/2$, where the annihilation proceeds through resonant Higgs exchange. 

Upon closer inspection, we find two viable regions of parameter space, see the panel displaying the allowed region in the Higgs portal coupling and the scalar DM fraction $R$. In one region, $\lambda_{HS}$ is of order ${\cal O}(10^{-2})$ and $R <1$, so that an additional DM component is required. In the second region,  the scalar particle constitutes a significant fraction or even all of DM, $R\lesssim 1$, but the Higgs portal coupling must be very small, of order  ${\cal O}(10^{-3} - 10^{-4})$. These two regions are a result of an interplay between  the strong velocity
dependence of the annihilation cross section near the resonance and the non-trivial scaling of the CR and GCE signals and the relic density with the fraction $R$ of scalar DM.

The best-fit points as well as their $\chi^2$ values are listed in table~\ref{tab:res} for the two regions described above.
For comparison we also show the results for the fit where we leave out the CR likelihood (GCE+constraints)
or the GCE likelihood (CR+constraints). Within the Higgs portal model the observations are very well compatible with each other. 
However, the CR signal prefers a flux corresponding to a slightly larger annihilation 
cross section. In the joint fit the nuisance parameters $\rho_\odot$ and $\log J/J_{\text{nom}}$
leave enough freedom to accommodate both signals. In fact, for $\rho_\odot$ and $\log J/J_{\text{nom}}$
the fit prefers somewhat larger and smaller values, respectively, than the nominal ones
(cf. lower panels in figure~\ref{fig:HPrepr}).
Note that $\rho_\odot$ also effects the direct detection rate $\propto \rho_\odot \, R \,\lambda_{HS}^2$.
As compared to the fit of the GCE presented in~\cite{Cuoco:2016jqt}, the improved LUX
limits and, to a lesser extent, the larger DM density $\rho_\odot$ further constrain large values of $R$ in the first region where $\lambda_{HS}$ is of order ${\cal O}(10^{-2})$. 
Another difference to the results of~\cite{Cuoco:2016jqt} arises from the fact that the recent 
results from dwarf galaxies are less constraining and, in particular, are not in tension with the GCE anymore. 
This allows for larger $\sv$ and hence for a smaller value of $\log J/J_{\text{nom}}$ while still fitting
the GCE signal. 
Note also that the $\rho_\odot$ range preferred by the Higgs portal model, 
$\rho_\odot= 0.6\pm0.1$ GeV/cm$^3$ (see figure~\ref{fig:HPrepr}),
is different from the case of the single channel fits where, instead, $\rho_\odot= 0.55\pm0.15$ GeV/cm$^3$.
The specific parameter region preferred by the Higgs portal fit, thus, further pushes $\rho_\odot$
toward higher values with respect to the single channel fit case.

 \begin{sidewaystable}[!h]
\begin{center}
\renewcommand{\arraystretch}{1.5}
\begin{tabular}{c | c  c  c | c c c } 
 & \multicolumn{3}{c |}{region 1} & \multicolumn{3}{c}{region 2}\\
 \hline
$\log L$ contribution  & 
GCE+constr.  & CR+constr. & GCE+CR+constr. &
GCE+constr.  & CR+constr. & GCE+CR+constr. \\ \hline\hline 
$m_{S}$\,{[}GeV{]}  & 
$62.58_{-0.04}^{+0.76}$  & 
$62.60_{-0.06}^{+0.21}$ & 
$62.58_{-0.03}^{+0.18}$&
$62.541_{-0.016}^{+0.003}$  & 
$62.532_{-0.009}^{+0.012}$ & 
$62.533_{-0.011}^{+0.011}$\\
$\lambda_{HS}$ & 
$0.017_{-0.003}^{+0.015}$  & 
$0.015_{-0.002}^{+0.006}$  & 
$0.015_{-0.001}^{+0.004}$ &
$0.0016_{-0.0013}^{+0.0060}$  & 
$0.00032_{-0.00012}^{+0.00815}$  & 
$0.00039_{-0.00017}^{+0.00561}$\\
$R$ & 
$0.019_{-0.018}^{+0.204}$  & 
$0.041_{-0.040}^{+0.124}$  & 
$0.020_{-0.018}^{+0.100}$ &
$0.021_{-0.019}^{+0.979}$  & 
$0.39_{-0.38}^{+0.61}$  & 
$0.29_{-0.28}^{+0.71}$\\
$\log J/J_{\text{nom}}$ & 
$-0.065_{-0.295}^{+0.341}$  & 
$-0.280_{-0.793}^{+0.351}$  & 
$-0.303_{-0.205}^{+0.304}$ &
$-0.099_{-0.275}^{+0.377}$  & 
$-0.415_{-0.590}^{+0.468}$  & 
$-0.316_{-0.201}^{+0.238}$\\
$\rho_\odot\,[\text{GeV}\text{cm}^{-3}]$  & 
$0.43_{-0.15}^{+0.15}$  & 
$0.56_{-0.08}^{+0.09}$ & 
$0.59_{-0.05}^{+0.1}$ &
$0.43_{-0.15}^{+0.15}$  & 
$0.56_{-0.09}^{+0.09}$ & 
$0.59_{-0.06}^{+0.09}$\\
$\langle\sigma v\rangle R^2\,[10^{-26}\, \text{cm}^{3}/\text{s}]$  & 
$1.36_{-0.44}^{+0.45}$  & 
$1.89_{-0.53}^{+0.72}$ & 
$1.73_{-0.47}^{+0.38}$ &
$1.36_{-0.45}^{+0.46}$  & 
$1.87_{-0.51}^{+0.72}$ & 
$1.70_{-0.32}^{+0.39}$\\

\hline
$\chi_{\text{GCE}}^{2}$  & 
26.22 & 
26.49 & 
26.69 &
26.47 & 
27.35 & 
26.88\\
$\chi_{\text{CR}}^2$  & 
52.32 & 
48.08 & 
48.42 &
57.14 & 
48.07 & 
48.42\\
\end{tabular}
\renewcommand{\arraystretch}{1}
\end{center}
\caption{
Fit parameters and $\langle\sigma v\rangle R^2$ for the best fit points of region 1 and 2 
taking into account the log-likelihood contributions from GCE+constraints, CR+constraints and CR+GCE+constraints.
Given errors are 1$\sigma$ uncertainties. We also show the corresponding $\chi^2_\text{GCE}$ and 
the $\chi^2_\text{CR}$.
}
\label{tab:res}
\end{sidewaystable}

%===================================================================
\section{Conclusion}\label{sec:summary} 
%===================================================================

In this paper  we analyze  antiproton data from AMS-02 searching for a signature of  DM annihilation.
Using the same methodology of \cite{Cuoco:2016eej}, we take into
account CR propagation uncertainties by fitting at the same time DM and propagation parameters.
With respect to \cite{Cuoco:2016eej} we explore a wider class of annihilation channels
including $gg$, $b \bar b$, $WW^{*}$, $ZZ^{*}$, $hh$ and $t\bar t$.
We find that almost all the channels provide similar hints of a DM annihilation at about 4$\sigma$ level (considering statistical uncertainties only)
with masses ranging from 40 and 130\,GeV depending on the annihilation channel.
Annihilation into $t\bar t$ provides a smaller fit improvement, at the 3$\sigma$ level.

We then investigate the compatibility of the antiproton DM hint with the
GCE performing a joint gamma-ray and antiproton fit where we further introduce two nuisance parameters
related to the distribution of DM in the vicinity of the Galactic center and in the Solar local neighborhood.
We find that the two signals are well compatible for most of the channels,
except for $gg$, where  the two are somewhat in tension.
Overall, we find that $b \bar b$,  $ZZ^{*}$ and $hh$ provides good fits to both the GCE and antiprotons, followed
by $WW^{*}$, which fits only slightly worse. 
$gg$ and $t\bar t$  are less favored, either because they do not fit well one of the two signals or because
the two signals are found to be in tension.
We also include in the fit the latest results from the analysis of dwarf galaxies in gamma rays 
and we find that dwarf constraints are compatible with the joint GCE and antiproton fit and 
do not change significantly the conclusions.

Finally, as an example, we perform the above joint fit for the specific case of the Higgs portal DM model,
including, in this case, also constraints from direct detection and collider searches.
We find that a surviving, although fine tuned, region corresponding to DM of mass equal to about 
$m_h/2$ annihilating via resonant Higgs exchange
satisfies all constrains and provides a good fit to both antiprotons and gamma rays.

%===================================================================
\section*{Acknowledgements}
%===================================================================

We acknowledge support by the German Research Foundation DFG through the 
research unit ``New physics at the LHC''.

\bibliographystyle{JHEP}
\bibliography{bibliography}

\end{document}